\documentclass{emulateapj}

\newcommand{\gsim}{\gtrsim}
\newcommand{\lsim}{\lesssim}

\def\micron {\ensuremath{\mu\mbox{{m}}}}
\newcommand{\um}{\micron}

\newcommand{\ks}{\ensuremath{K_{\rm s}}}

\newcommand{\pasa}{PASA}

\newcommand{\gcrt}{GCRT~J1745$-$3009}

\shorttitle{A Near-IR Search for \gcrt}
\shortauthors{Kaplan et al.}

\begin{document}

\title{A Search for the Near-Infrared Counterpart to \gcrt\altaffilmark{1,2}}

\author{D.~L.~Kaplan\altaffilmark{3,4}, S.~D.~Hyman\altaffilmark{5},
S.~Roy\altaffilmark{6}, R.~M.~Bandyopadhyay\altaffilmark{7},
D.~Chakrabarty\altaffilmark{4}, N.~E.~Kassim\altaffilmark{8},
T.~J.~W.~Lazio\altaffilmark{8}, P.~S.~Ray\altaffilmark{9}}
\slugcomment{Accepted for publication in ApJ}
\altaffiltext{1}{This paper includes data gathered with the 6.5 meter
  Magellan Telescopes located at Las Campanas Observatory, Chile.}
\altaffiltext{2}{Based on observations obtained at the Gemini Observatory, which is operated by the
Association of Universities for Research in Astronomy, Inc., under a cooperative agreement
with the NSF on behalf of the Gemini partnership: the National Science Foundation (United
States), the Science and Technology Facilities Council (United Kingdom), the
National Research Council (Canada), CONICYT (Chile), the Australian Research Council
(Australia), Minist\'{e}rio da Ci\^{e}ncia e Tecnologia (Brazil) and
SECYT (Argentina).}

\altaffiltext{3}{Pappalardo Fellow and Hubble Fellow}
\altaffiltext{4}{Kavli Institute for Astrophysics and Space Research
and Department of Physics, Massachusetts Institute of Technology,
Cambridge, MA 02139; dlk@space.mit.edu, deepto@space.mit.edu.}
\altaffiltext{5}{Department of Physics and Engineering, Sweet Briar
College, Sweet Briar, VA 24595; shyman@sbc.edu.}
\altaffiltext{6}{NCRA-TIFR, Pune University Campus, Pune-7, India; roy@ncra.tifr.res.in.} 
\altaffiltext{7}{Department of Astronomy, University of
Florida, Gainesville, FL 32611; reba@astro.ufl.edu.}
\altaffiltext{8}{Remote Sensing Division, Naval Research Laboratory,
Washington, DC 20375-5351; namir.kassim@nrl.navy.mil, joseph.lazio@nrl.navy.mil}
\altaffiltext{9}{Space Science Division, Naval
  Research Laboratory, Washington, DC 20375-5352;
  Paul.Ray@nrl.navy.mil.}

\begin{abstract}
We present an optical/near-infrared search for a counterpart to the
perplexing radio transient \gcrt, a source located $\sim1\degr$ from
the Galactic Center.  Motivated by some similarities to radio bursts
from nearby ultracool dwarfs, and by a distance upper limit of
$70\,$pc for the emission to not violate the $10^{12}\,$K brightness
temperature limit for incoherent radiation, we searched for a nearby
star at the position of \gcrt.  We found only a single marginal
candidate, limiting the presence of any late-type star to $>1\,$kpc
(spectral types earlier than M9), $>200\,$pc (spectral types L and
T0--T4), and $>100\,$pc (spectral types T4--T7), thus severely
restricting the possible local counterparts to \gcrt.  We also exclude
any white dwarf within $1\,$kpc or a supergiant star out to the
distance of the Galactic Center as possible counterparts.  This
implies that \gcrt\ likely requires a coherent emission process,
although whether or not it reflects a new class of sources is unclear.
\end{abstract}



\section{Introduction}
A blind search program using the Very Large Array (VLA) at 330$\,$MHz ($90\,$cm),
made possible by new wide-field radio imaging techniques, has resulted
in the discovery of two radio transients near the Galactic
Center. One of these, \gcrt, exhibits dramatic 1-Jy 
outbursts that last about 10$\,$min and have a recurrence interval of
about 77$\,$minutes \citep{hlk+05}. 


\gcrt\ exhibited 5 bursts at 330$\,$MHz, each lasting about 10$\,$minutes,
at apparently regular intervals of 77.1$\,$min during a 7-hour VLA
observation in September 2002.  After several non-detections during
the summer of 2003, a single burst was detected during a short Giant
Metrewave Radio Telescope (GMRT) observation in September 2003
\citep{hlr+06}, confirming that it is  recurrent.  It has
since also been faintly detected in an additional GMRT observations from 2004,
where it had a very steep nonthermal radio spectrum across the
$32\,$MHz bandpass
\citep{hrp+07}.  It is at Galactic coordinates $\ell = 358.89\degr$,
$b=-0.54\degr$, so it may be as close as 180$\,$pc to the Galactic
Center (GC), though the distance of \gcrt\ from the Earth is completely unknown.  If it is at a
distance $d>70\,$pc from Earth, then the radio flux density combined
with the rapid ($\sim 2\,$min) decay of the bursts constrains its
brightness temperature to exceed the $10^{12}\,$K limit for incoherent
synchrotron radiation \citep{readhead94} and thus is very likely a
coherent emitter.  The positional coincidence with the GC, along with
the greatly increasing source density near the Center, suggests it is
at a distance of about 8$\,$kpc, but this could be a selection effect
because all the fields searched for transients in this program were in
that direction.  Consequently, it is very important to rule out local
classes of sources as possible explanations.

%

Among the many possibilities discussed for the origin of \gcrt,
several involved low-mass stars or substellar objects (brown dwarfs,
or BDs), which we generically call ultracool dwarfs.  These were
initially rejected as being unlikely or inconsistent with the observations
\citep{hlk+05}.  However, additional observations are required to
strengthen the case against the ultracool dwarfs and to confirm that
the radio source is indeed a new type of source.  Alternative models
have also been proposed, ranging from a nulling pulsar \citep{kp05}, a
double pulsar \citep*{tpt05}, a transient white dwarf (WD) pulsar
\citep{zg05}, to a precessing radio pulsar \citep{zx06}, but these
will largely be constrained through other observations.

A number of BDs and late-type stars have now been detected at GHz
frequencies, in many cases despite the absence of flaring optical or
X-ray emission that would accompany radio emission from normal stars
(see \citealt{berger06} for a review).  These objects exhibit a range
of nonthermal flaring behaviors, from strong, narrowband, fully
polarized, bursting emission with frequency drifts \citep{bp05}, to
periodic bursts suggestive of pulsar-like beaming \citep{hbl+07}, to
order-of-magnitude variations spanning years \citep{adh+07}.  In
particular, the behavior found by \citet{hbl+07} for the M9 dwarf
TVLM~513$-$46546 is reminiscent of \gcrt: bright bursts of coherent
emission that follow the 2~hr rotation of the star.  Although the
bursts are orders of magnitude weaker than those seen for \gcrt, the
observations differ by an order of magnitude in wavelength, and there
are no current limits on the radio behavior of ultracool dwarfs at
meter wavelengths.

We then have two independent but complementary motivations for
searching for a counterpart to \gcrt\ in the near-infrared (near-IR)
and optical bands.  First, we wish to see if there is {any} possible
counterpart closer than $70\,$pc: otherwise, we must conclude that the
emission from \gcrt\ is coherent.  Second, given the similarity
between it and TVLM~513$-$46546, we wish to see if \gcrt\ is
associated with a nearby ultracool dwarf, although such an object
could be more distant than $70\,$pc as they can emit coherently.  A
single ultracool dwarf in the error circle of \gcrt\ would be strong
evidence that it is the counterpart, even in the absence of a precise
position, as one expects $<1$ such object in a $2\arcmin$ field
(\citealt{rkl+99}; \citealt*{cbk08}).  We concentrated on the near-IR
for several reasons: (1) if the source is at the distance of the
Galactic Center, the average $K$-band extinction, $A_K$, at this
position is approximately 2--3 magnitudes, corresponding to a visual
extinction of $\geq$20 magnitudes (see e.g., \citealt{dsbb03}); (2) if
the source is a relatively nearby late-type star, it would be
intrinsically red; and (3) the field towards the GC is very crowded,
and near-IR observations often have better seeing than bluer bands.
We augmented the traditional near-IR bands ($JH\ks$) with $I$-band to
aid in color discrimination (see below).

The structure of this paper is as follows.  First, to aid the analysis
that follows, we update our determination of the radio position of
\gcrt\ in \S~\ref{sec:position}.  We then describe our optical and
near-infrared data in \S~\ref{sec:obs}.  The primary analysis is in
\S~\ref{sec:bd}, where we attempt to constrain any ultracool dwarf
counterpart to \gcrt.  Following this, in \S~\ref{sec:other} we also
examine what constraints on other types of counterparts we can
determine from our data.  Finally, we give our conclusions in
\S~\ref{sec:conc}.

\section{An Updated Radio Position}
\label{sec:position}
Before we search for counterparts at other wavelengths, we need to use
the most accurate position possible for \gcrt, and we must correct for
the significant ionospheric refraction prevalent at low frequencies.
We therefore present a revised analysis of the radio position of
\gcrt.  This position was obtained by registering the 330-MHz 2003 and
2004 GMRT images with respect to the revised 6- and 20-cm source
positions determined by \citet*{wbh05} in their reanalysis of the
Galactic plane radio survey (GPRS) of 586 compact sources
\citep{zhb+90,hzbw92,bwhz94}. Our previous reports on \gcrt\ used
either the initial GPRS survey or the much lower resolution NRAO VLA
sky survey \citep{ccg+98} to register the 330-MHz images.

We compared 13 and 19 sources within the primary beams of the 2003 and
2004 observations, respectively, with their 5- and 1.4-GHz (6- and
20-cm) counterparts. Five of the sources were detected at both 5 and
1.4$\,$cm, and the average source positions were calculated for
these. The 6 or 20$\,$cm coordinates of the other 27 sources were
shifted by one-half of the average difference ($0.018^{\rm s}\pm
0.059^{\rm s}$ in right ascension and $0.64\arcsec \pm 0.81\arcsec$ in
declination) we found between the 5- and 1.4-GHz positions of the five
sources mentioned above and three others that were not detected at
330$\,$MHz.

\begin{figure*}
\plotone{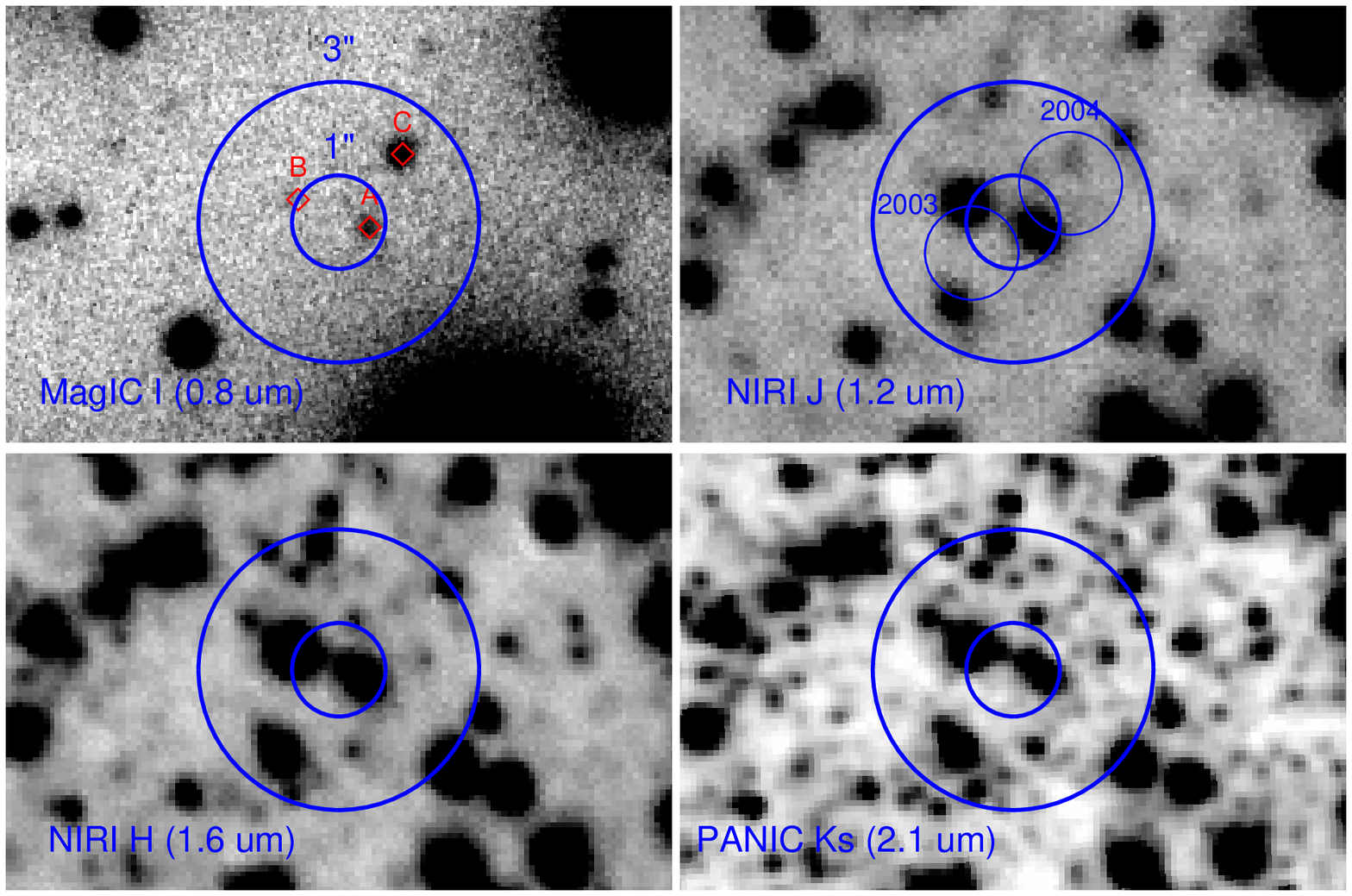}
\caption{Images of the field of \gcrt.  We show our best images in $I$ (top
  left), $J$ (top right), $H$ (bottom left) and $\ks$ (bottom right)
  bands.  We give the individual error circles from the 2003 and 2004
  observations ($J$-band only), as well as the weighted average.  The error circles on
  the average position have radii of $1\arcsec$ and $3\arcsec$,
  corresponding to 1- and 3-$\sigma$ uncertainties.  North is up, and
  East to the left.  The 3 objects with $I$-band detections from
  Table~\ref{tab:phot} are labeled in the $I$-band image.
}
\label{fig:img}
\end{figure*}

The position correction is determined to be $0.185^{\rm
s}\pm0.071^{\rm s}$ in R.A.\ and $-2.87\arcsec\pm 0.92\arcsec$ in
Dec.\ for the 2003 observation and $0.013^{\rm s}\pm0.052^{\rm s}$ in
R.A.\ and $-2.88\arcsec\pm0.77\arcsec$ in Dec.\ for the 2004
observation. The measured position of \gcrt\ was corrected by these
differences at each of the two epochs.  After correction, the two
positions are consistent at roughly the 1$\,\sigma$ level, as shown in
Figure~\ref{fig:img}.  A weighted average of the two yields a position
of (J2000) $17^{\rm h}45^{\rm m}05\fs15$,
$-30\degr09\arcmin52\farcs7$, with a 1-$\sigma$ uncertainty of $\pm
0.7\arcsec$ on each coordinate.  To be conservative and allow for
astrometric frame uncertainties (which should be $\approx 0.2\arcsec$
in each coordinate; see below), we use a position uncertainty in the near-IR of
$1\arcsec$, and consider objects within 3-$\sigma$ ($3\arcsec$ radius)
of the position of \gcrt.

\begin{deluxetable*}{l l c c c c}
\tablecaption{Observation Summary\label{tab:obs}}
\tablewidth{0pt}
\tablehead{
\colhead{Date} & \colhead{Telescope} & \colhead{Instrument} &
\colhead{Band} & \colhead{Exposure} & \colhead{Seeing} \\
& & & & \colhead{(sec)} & \colhead{(arcsec)} \\
}
\startdata
2005-Jul-12 & Magellan I/Baade & PANIC & $\ks$ (2.1$\,$\um) & 1800 & 0.4\\
2005-Jul-13 & & & $J$ (1.2$\,$\um) & 1920 & 0.8\\
2005-Jul-14 & & & $H$ (1.6$\,$\um) & 2460 & 0.7\\
2005-Jul-14 & Gemini North & NIRI & $J$ (1.2$\,$\um) & \phn840 & 0.5\\
            & & & $H$ (1.6$\,$\um) & \phn840 & 0.5 \\
 & & & $\ks$ (2.1$\,$\um) & \phn780 & 0.4 \\
2006-Jun-20 & Magellan II/Clay & MagIC & $I$ (0.8$\,$\um) & 3660 & 0.4\\
\enddata
\tablecomments{PANIC is  Persson's Auxiliary Nasmyth Infrared Camera on
  the 6.5-m Baade (Magellan I) telescope \citep{mpm+04}.  
NIRI is the Near-Infrared Imager on the 8-m Gemini North telescope
\citep{hji+03}. MagIC is the  Raymond and Beverly 
  Sackler Magellan Instant Camera on the 6.5-m Clay (Magellan II). 
}
\end{deluxetable*}

\section{Optical and Infrared Observations \& Reduction}
\label{sec:obs}
We obtained near-infrared ($JH\ks$ bands, covering wavelengths
1.2--2.1$\,$\um) photometric observations of \gcrt\ in the summer of 2005
with both PANIC on the Magellan~I/Baade telescope and NIRI on the
Gemini North telescope: see Table~\ref{tab:obs} for details.  We also
observed \gcrt\ in the $I$-band (0.8$\,$\um) in 2006 with Magellan.  

The near-infrared (PANIC and NIRI) reduction was similar for both
instruments.  For NIRI, there is an \texttt{IRAF} package to reduce
the data available from the NIRI web site\footnote{See
\url{http://www.us-gemini.noao.edu/sciops/instruments/niri/NIRIIndex.html}.}.
We proceeded through the steps of this package, flatfielding the data,
subtracting the sky, shifting the images, and adding them together.
For PANIC, we used our own routines in \texttt{PyRAF}, but the steps
were similar.

We referenced the astrometry and photometry of the PANIC data to
the Two Micron All Sky Survey (2MASS; \citealt{2mass}).  For the astrometry, we used 210 2MASS stars that we identified
as not blended or badly saturated for $J$ and $H$ bands and 135 stars
for $\ks$-band, getting rms residuals of $0\farcs11$ in each
coordinate. The net uncertainty is $\approx 0.2\arcsec$ in each
coordinate, dominated by the $\approx 0.15\arcsec$ uncertainty in the
tie between 2MASS and the International Celestial Reference System (ICRS).
We used fewer stars for photometry, since we avoided
sources that showed any signs of saturation or significant
non-linearity in the data.  However, we still used $>30$ stars in
$\ks$-band and $>100$ stars in $J$ and $H$ bands.  We estimate
zero-point uncertainties of 0.1~mag for $\ks$-band and 0.05~mag for
$J$ and $H$ bands, with the larger uncertainty at $\ks$-band from the
smaller number of stars and the greater effects of saturation.

For the NIRI data, we used 116 2MASS stars for the astrometry, and obtained
rms residuals of $0\farcs09$ in each coordinate.  For the photometry,
we tried to reference it to 2MASS and found zero-point magnitudes
reasonably close to those in the NIRI manual, but many of the 2MASS
stars showed some signs of saturation and therefore we wanted to check
with another method of calibration.  We used an observation of the
standard star FS~140.  We corrected for the difference in airmass
between the standard star observation and the observations of \gcrt\
using the extinction coefficients listed for the Keck
telescope\footnote{See
\url{http://www.us-gemini.noao.edu/sciops/instruments/niri/standards/UKIRT-fs.html}.}
(also on Mauna~Kea), although these corrections were minor.  There
might have been some small variations in extinction due to clouds over
the course of the night (from the Canada France Hawaii Telescope sky
probe).  Overall, the zero-point magnitudes from both methods agreed
to within 0.1~mag, and measurements of stars from PANIC and NIRI
agreed to within this limit as well.  We therefore assign a zero-point
uncertainty of 0.1~mag to all of the NIRI data.

For the $I$-band data, we did standard reduction in \texttt{IRAF} by
subtracting overscan regions, merging the data from four amplifiers,
and flatfielding the data with twilight flats.  We registered the
astrometry to that of the PANIC $\ks$-band image, using 248 stars that
were not blended or badly saturated, and obtained rms residuals of
$0\farcs07$ in each coordinate.  For photometry, we used observations
of the standard fields L113-339 and NGC~6093 \citep{l92,s00}, and we
estimate a zero-point uncertainty of 0.1~mag.

\section{Search for an Ultracool Dwarf}
\label{sec:bd}
When we consider seeing and depth, our best set of images in the
near-IR are the NIRI images in $J$ and $H$ bands, and the PANIC
$\ks$-band image: unless otherwise noted, all references to $J$, $H$,
and $\ks$ band images refer to those.  We show the best near-IR images
plus the $I$-band image in Figure~\ref{fig:img}.  The other near-IR
data are mostly useful as a check, but we note that we see no
variability in the sources within the error circle over the 1--2~day
span between the observations.

Since the $\ks$-band image shows by far the most objects and has the
best seeing, we computed aperture photometry for all of the sources in
that image using \texttt{sextractor} \citep{ba96}.  We then used those
source positions to run \texttt{sextractor} on the $I$, $J$ and $H$
images, generating a set of source photometry in four bands, which we
supplemented with 2MASS photometry for the sources that were saturated
on our images.  We repeated this procedure but starting this time from
the $I$-band image: this picked up many fewer sources, but they sample
a larger range in color space.

Overall, we found 3-$\sigma$ limiting magnitudes for the
$I,J,H,\ks$-band images of 26, 21, 20, and 19, respectively.  However,
these limits apply over the images as a whole, and especially for the
longer wavelengths are dominated by the effects of confusion.
Therefore, we can detect $\ks$ sources that are considerably fainter
than the quoted limit, although not in all regions.

For the sources in the 3-$\sigma$ error circle we were more careful
about the photometry.  We used PSF fitting routines from
\texttt{DAOPHOT} \citep{stetson87} within \texttt{IRAF}.  For each
band, we modeled the PSF from an ensemble of about 50 stars that were
well separated from all neighbors and where our aperture photometry
was not corrupted by any bad pixels or cosmic rays.  We then
iteratively fit the PSF models to the sources within the error circle,
identifying the brighter objects, subtracting them, and then examining
the residual image for any fainter neighbors.  What is most striking
about the images is the crowding at $\ks$-band compared to the
relatively empty field at $I$-band: we detect at least $26$~sources in
the former, compared to only 3 in the latter.

\begin{figure*}
\plotone{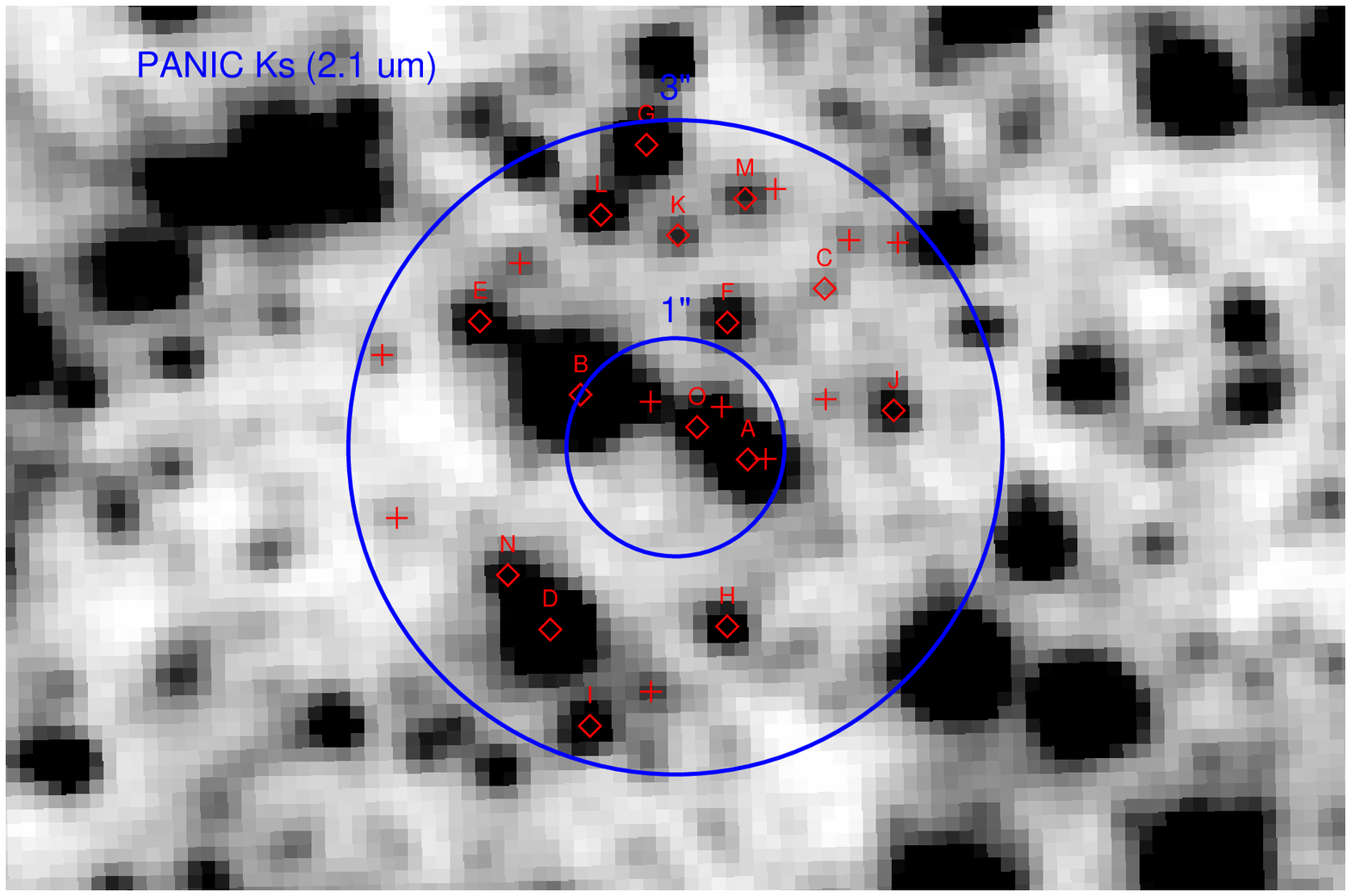}
\caption{PANIC \ks\ image of the field of \gcrt.  Again, the error
   circles have radii of $1\arcsec$ and $3\arcsec$, corresponding to 1-
   and 3-$\sigma$ uncertainties.  We label and put diamonds on all of the sources with
   detections in at least two bands (Table~\ref{tab:phot}), and put
   crosses on the remaining objects with only $\ks$-band detections.  }
\label{fig:imglabel}
\end{figure*}

For the 15 objects with  detections in at least two bands, we give the PSF
photometry in Table~\ref{tab:phot} (also see Figure~\ref{fig:imglabel}), where we have attempted to
combine the photometry from the different images even though it is not
always perfectly clear how to do so given the differences in seeing
and brightness.  For instance, in the $\ks$-band image object A is
relatively bright, but is surrounded by a number of nearby objects
that are $>1\,$mag fainter.  However, in the $J$-band image all of
those objects are blended together, and so we ascribe all of the flux
to object A in that band.  It is clear that we have detected sources
fainter than the nominal limiting magnitudes given above.  There may
be some additional faint objects remaining in the $\ks$-band image
that we could have included, but none of these is detected in any
other band, so they are of limited utility in our attempt to find a
counterpart to \gcrt.

To define the sort of object that we are searching for, we consider
the results of \citet{rbc+01}, \citet{dhv+02}, and \citet{vhl+04}.
Together, these papers discuss the classification and absolute
photometry of cool dwarf stars and BDs.  They establish the broad-band
colors as a function of spectral type for stars with spectral types
ranging from late M through T (also see \citealt{cbk08}, which draws
data from many of the same sources and gives  distance limits
similar to those presented here).  In particular, \citet{dhv+02} includes
photometry in the $IJH\ks$ bands.  We note, though, that there may be
difficulties in comparing our photometry with results from the
literature, especially in the $I$-band.  That is because, as is
discussed by \citet{dhv+02}, the majority of the flux from cool stars
in that bandpass is at the very red edge, and therefore details of the
filter cutoff and the detector response become quite important.  When
calibrated, as we did, with an ensemble of stars of a range of
moderate spectral types, the calibration may have errors when applied
to very red objects.  We did not have sufficient calibration data to
solve for color or extinction terms.  We must therefore allow for
additional zero-point uncertainties when examining potential BDs in
the $I$-band; this should be less than a few tenths of a magnitude for
stars earlier than L5, but for later stars could even be $>1\,$mag
(A.~Burgasser, priv.\ communication).  There are also differences
between the near-IR filters used (CIT vs.\ 2MASS, usually), but those
are generally smaller than our zero-point uncertainties and we have
applied basic corrections for them \citep{c01,dhv+02}.

\begin{figure}
\plotone{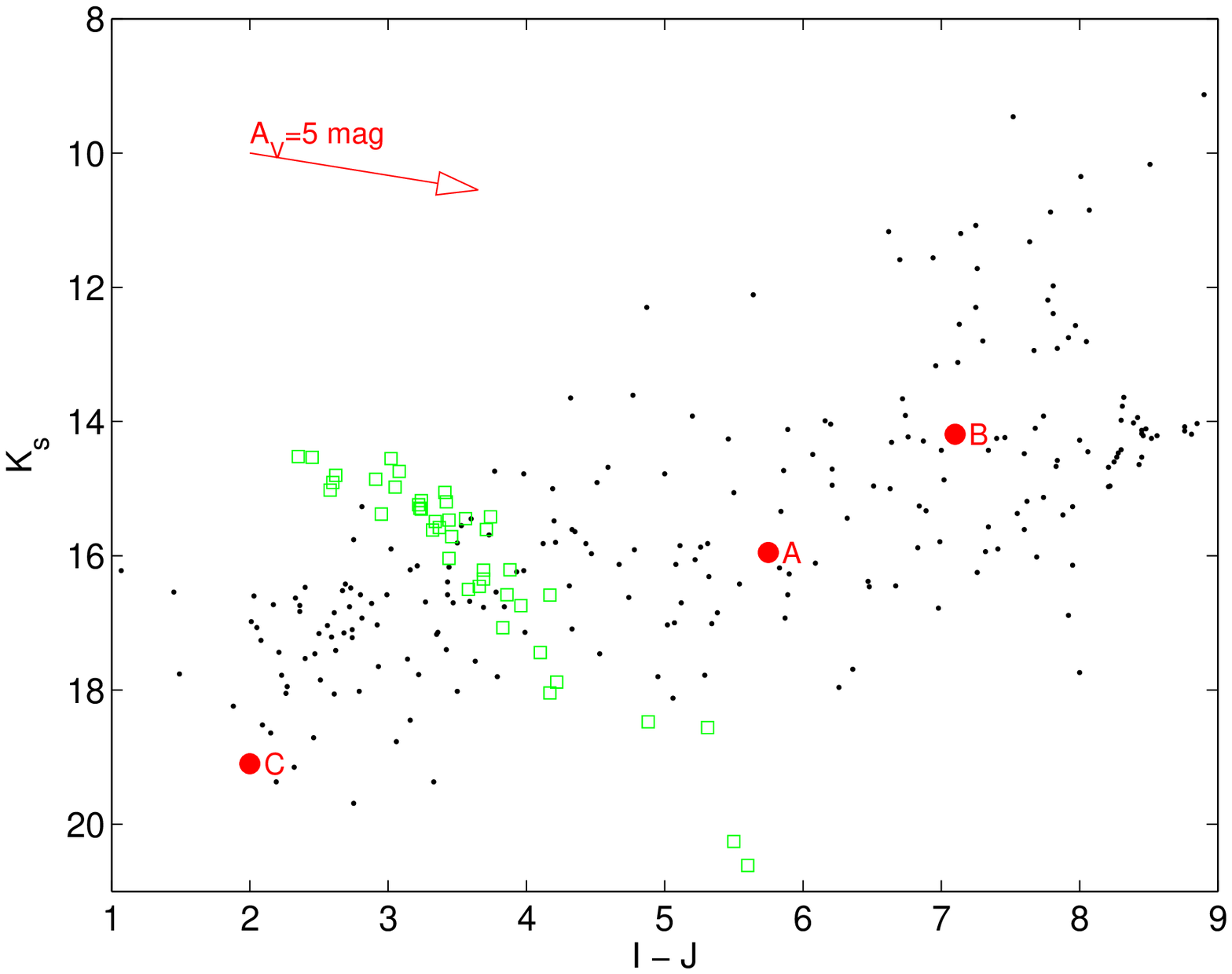}
\plotone{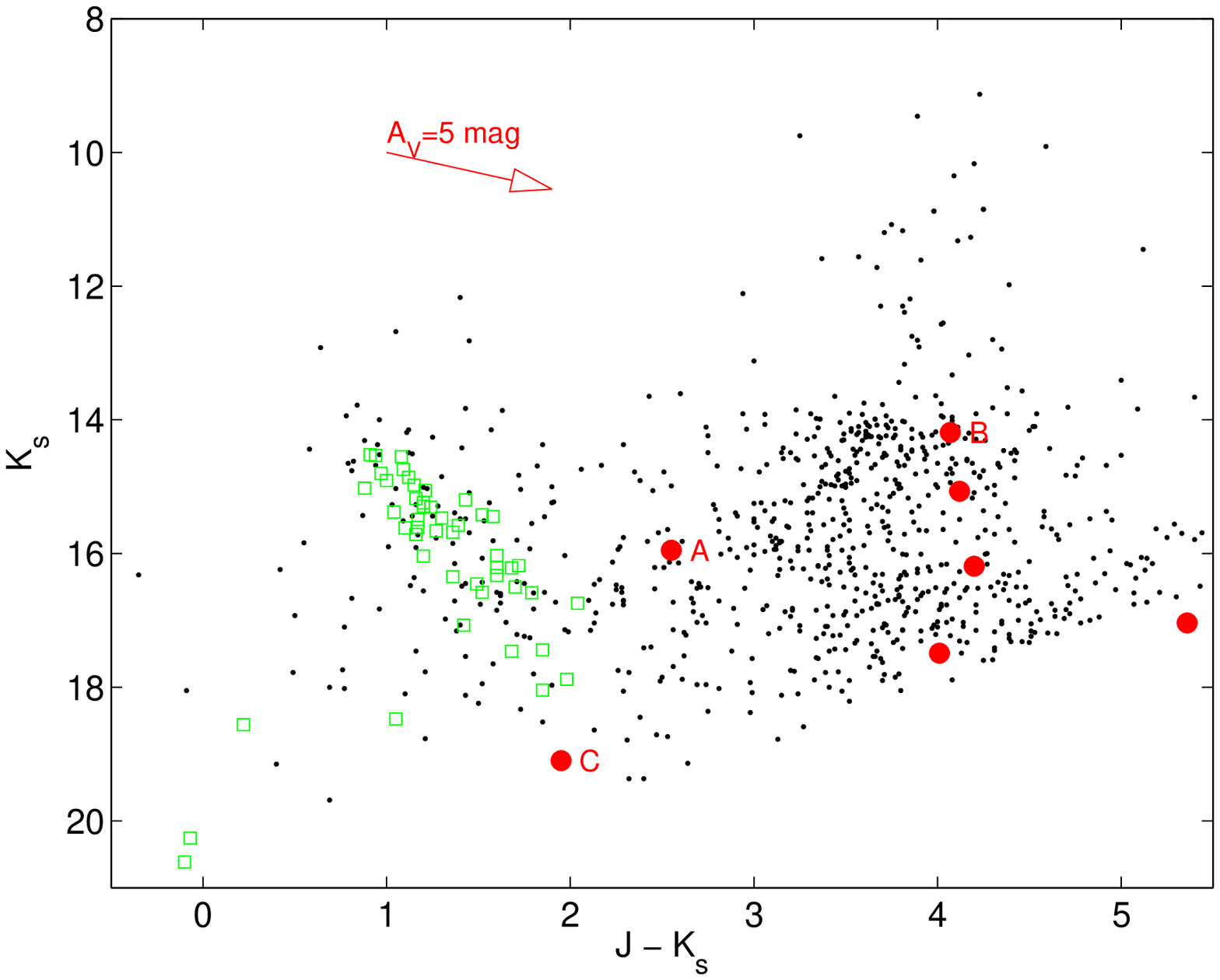}
\plotone{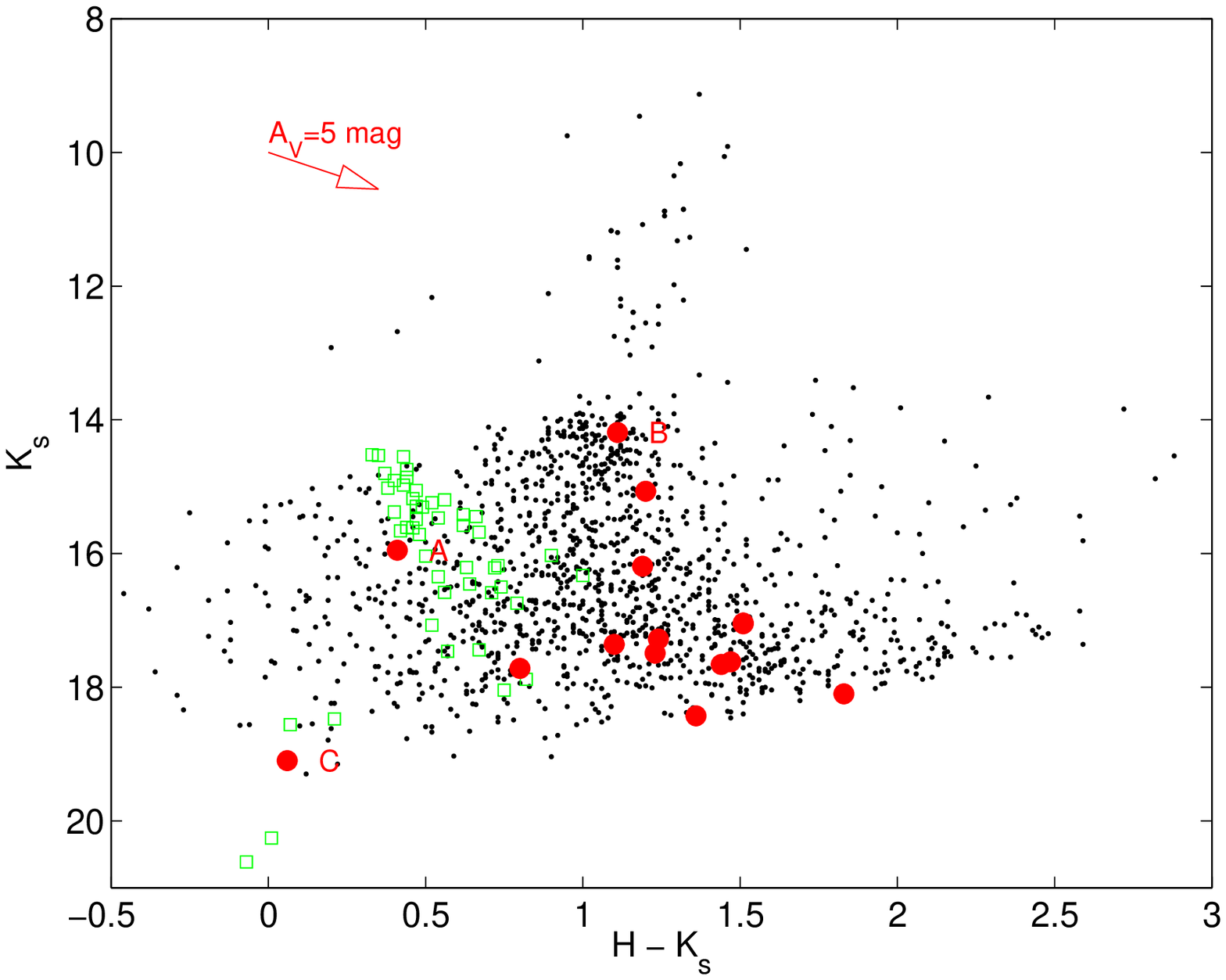}
\caption{Color-magnitude diagrams for the field of \gcrt.  We plot
  $I-J$ (\textit{top}), $J-\ks$ (\textit{middle}), and $H-\ks$
  (\textit{bottom}) color versus \ks\ magnitude.  The points are
  objects from the field, the red filled circles are the objects from the
  3-$\sigma$ error circle (with the objects that have $I$-band
  detections labeled according to Tab.~\ref{tab:phot}) and the green
  squares are the objects from \citet{dhv+02} shifted to a distance of
  100~pc.  Reddening vectors for $A_V=5\,$mag are also plotted.}
\label{fig:cmd}
\end{figure}

Objects with spectral types down to late L have absolute $I$
magnitudes of $<19$, absolute $J$ magnitudes of $<15$, and $\ks$
magnitudes of $<13$ or so.  The $I-J$ color for those objects ranges
from $\approx 2.5$ at the bright end to $\approx 4$ at the faint end,
while $J-\ks$ goes from 1 to 2 over the same range.  For later-type
stars, the absolute $\ks$ magnitude increases to about 16.5 at T9
($M_I$ increases to about 22 and $M_J$ increases to about 17), and the
stars become redder in $I-J$ (up to about 5.5), but they become bluer
in $J-\ks$ (to about 0).     See Figure~\ref{fig:cmd}.

\begin{deluxetable*}{c c c c c c c}
\tablecaption{Objects in the \gcrt\ $3\arcsec$ Error Circle With
  Detections in At Least Two Bands\label{tab:phot}}
\tablewidth{0pt}
\tablehead{
\colhead{ID} & \colhead{$\alpha_{\rm J2000}$} & \colhead{$\delta_{\rm
    J2000}$} & \colhead{$I$} & \colhead{$J$} & \colhead{$H$} &
\colhead{$\ks$} \\
}
\startdata
A\tablenotemark{a} & $17^{\rm h}45^{\rm m}05\fs09$ & $-30\degr09\arcmin52\farcs8$ & 24.25(4)\phd & 18.50(2)\phd & 16.36(2)\phd & 15.95(6)\phd  \\
B & $17^{\rm h}45^{\rm m}05\fs21$ & $-30\degr09\arcmin52\farcs2$ & 25.36(10) & 18.26(2)\phd & 15.30(1)\phd & 14.19(3)\phd  \\
C & $17^{\rm h}45^{\rm m}05\fs04$ & $-30\degr09\arcmin51\farcs3$ & 23.05(1)\phd & 21.05(7)\phd & 19.16(8)\phd & 19.10(15)  \\
D & $17^{\rm h}45^{\rm m}05\fs23$ & $-30\degr09\arcmin54\farcs4$ & $\ldots$ & 19.19(3)\phd & 16.27(1)\phd & 15.07(3)\phd  \\
E & $17^{\rm h}45^{\rm m}05\fs28$ & $-30\degr09\arcmin51\farcs6$ & $\ldots$ & 22.40(24) & 18.55(7)\phd & 17.04(5)\phd  \\
F & $17^{\rm h}45^{\rm m}05\fs11$ & $-30\degr09\arcmin51\farcs6$ & $\ldots$ & 21.50(9)\phd & 18.72(5)\phd & 17.49(6)\phd  \\
G & $17^{\rm h}45^{\rm m}05\fs17$ & $-30\degr09\arcmin50\farcs0$ & $\ldots$ & 20.39(6)\phd & 17.38(3)\phd & 16.19(4)\phd  \\
H & $17^{\rm h}45^{\rm m}05\fs11$ & $-30\degr09\arcmin54\farcs4$ & $\ldots$ & $\ldots$ & 18.46(7)\phd & 17.36(5)\phd  \\
I & $17^{\rm h}45^{\rm m}05\fs21$ & $-30\degr09\arcmin55\farcs3$ & $\ldots$ & $\ldots$ & 18.56(5)\phd & 17.05(6)\phd  \\
J & $17^{\rm h}45^{\rm m}04\fs99$ & $-30\degr09\arcmin52\farcs4$ & $\ldots$ & $\ldots$ & 19.10(9)\phd & 17.66(7)\phd  \\
K & $17^{\rm h}45^{\rm m}05\fs14$ & $-30\degr09\arcmin50\farcs8$ & $\ldots$ & $\ldots$ & 19.79(8)\phd & 18.43(6)\phd  \\
L & $17^{\rm h}45^{\rm m}05\fs20$ & $-30\degr09\arcmin50\farcs6$ & $\ldots$ & $\ldots$ & 18.52(4)\phd & 17.28(5)\phd  \\
M & $17^{\rm h}45^{\rm m}05\fs10$ & $-30\degr09\arcmin50\farcs4$ & $\ldots$ & $\ldots$ & 19.93(10) & 18.10(5)\phd  \\
N & $17^{\rm h}45^{\rm m}05\fs26$ & $-30\degr09\arcmin53\farcs9$ & $\ldots$ & $\ldots$ & 19.09(7)\phd & 17.62(5)\phd  \\
O\tablenotemark{a} & $17^{\rm h}45^{\rm m}05\fs13$ & $-30\degr09\arcmin52\farcs5$ & $\ldots$ & $\ldots$ & 18.52(7)\phd & 17.72(8)\phd  \\

\enddata
\tablenotetext{a}{These objects are  within  $1\arcsec$ of the radio
  position of \gcrt.}
\tablecomments{See Figure~\ref{fig:imglabel}.  The quantities in
  parentheses are 1-$\sigma $ statistical uncertainties on the last
  digit only: there are additional zero-point uncertainties of
  0.1~mag.  
} 
\end{deluxetable*}

\subsection{Implications of Non-detections}
\label{sec:nondetect}
In searching for a nearby, ultracool counterpart to \gcrt, we  first treat
all objects  detected in the error circle as unrelated background
stars (we will examine the validity of this assumption in
\S~\ref{sec:detect}).  This will enable us to make some general
statements about the field and to derive some useful relations.  We
assume that any extinction will be minimal: at the distances of
interest here ($\lsim 300\,$pc), we expect $A_V\lsim0.2\,$mag or so
(based on \citealt*{dcllc03}), which translates to an extinction of
$\lsim 0.02\,$mag at $\ks$-band.

We will start our examination with the $I$-band image: while there are
calibration uncertainties, the depth of the image and the lack of
crowding make it preferable.  The PSF fitting and subtraction did not
reveal any fainter objects in the error circle, and the background is
regular enough that we are not dominated by confusion.  Using the data
from \citet{dhv+02}, we derive the absolute $I$ magnitude as 
a function of spectral type $ST$:
\begin{equation}
M_I(M,L) \approx 10.5 + 0.45ST
\end{equation}
where $ST=7$ for spectral type M7, 10 for spectral type L0, and goes
up to 20 for spectral type T0; the dispersion is about 0.5~mag, and it
is not valid for later spectral types.  We use this with our $I$-band
detection threshold of $I\approx 26$ to limit the spectral type as a
function of distance $d$ that a non-detection in $I$-band implies:
\begin{equation}
ST_{{\rm non-det},I}(M,L) \gsim 23.4-11.1\log_{10}\left(\frac{d}{100\,{\rm pc}}\right).
\end{equation}
So for no detection in $I$-band, we can exclude almost  all M and
L type objects out to $200\,$pc.  Even with the scatter on the $M_I(ST)$
relation and the calibration uncertainties, this should be a
relatively robust result.  More typical main-sequence
stars (spectral types K and earlier) are excluded closer than about
$5\,$kpc (see discussion of object C in \S~\ref{sec:detect}).

Moving to $J$-band, we still detect only 7 objects.  Here the PSF
subtraction did reveal fainter objects, and we have included them to
the best of our ability.  With an
approximate upper limit of $J<21$ and the relation from
\citet{dhv+02}: $M_J(M,L)\approx8.38+0.341ST$,
we can again limit:
\begin{equation}
ST_{{\rm non-det},J}(M,L) \gsim 22.4 - 14.7\log_{10}\left(\frac{d}{100\,{\rm pc}}\right).
\end{equation}
so the implications are similar.  For later spectral types, we can use
the comparable relation from \citet{vhl+04} for T dwarfs, and we limit
possible objects to later than T7 or so at $100\,$pc and and later
than T4 or so at $200\,$pc (the relation is more complicated than the
linear forms above, so we do not give a simple expression).

TVLM~513$-$46546, the pulsating dwarf found by \citet{hbl+07}, is of
spectral type M9 ($ST=9$) and at a distance of $10.5\,$pc
\citep{lag+01}.  To have such an object be fainter than our $I$-band
limit would require a distance of $>2\,$kpc, although this will be
somewhat of an over estimate, since for $>0.5\,$kpc the extinction
will start to contribute.  Assuming an average extinction in the
$I$-band of $\sim 1\,{\rm mag}\,{\rm kpc}^{-1}$, we can still exclude
an M9 star within $1.3\,$kpc.  The observed flux density from
TVLM~513$-$46546 peaks at a few mJy at frequencies of 4 and 8$\,$GHz
(8 and 4$\,$cm).  In comparison, bursts from \gcrt\ have peak flux
densities of $\sim 1\,$Jy at 330$\,$MHz, although this has been
observed to vary by a factor of $\sim 10$ \citep{hrp+07}.  Scaling
TVLM~513$-$46546 to 1$\,$kpc we would expect peak flux densities of
$<1\,\mu$Jy at 5$\,$GHz (6$\,$cm); having the flux density be $\sim
1\,$Jy at $330\,$MHz would require an average spectral index of
$\alpha\sim -5$ over more than a decade of frequency (where
$S_{\nu}\propto \nu^{\alpha}$), inconsistent with typical spectral
indices observed from M dwarfs \citep[0 to $-1$;][]{gsbf93}.  Similar
conclusions come from comparison with the L3.5 dwarf
2MASS~J00361617+1821104 observed by \citet{brr+05}, which has a
$3\,$hr periodicity in the radio (implied spectral index $\sim -6$ for
a flux density of $\sim 1\,$Jy at $330\,$MHz).  This is far steeper
than almost any known source \citep[e.g.,][]{kccg00} and inconsistent
with the flat-spectrum bursts from TVLM~513$-$46546 \citep{had+06} and
dwarf stars in general \citep{gb96}, although such steep emission
should be possible \citep{erickson99} and the bursts for \gcrt\
reported by \citet{hrp+07} do have a very steep spectral index of
$\alpha=-13.5$ across a narrow (32$\,$MHz) bandpass.  \citet{bbdd90}
found a similarly steep spectrum  across a $\sim 40\,$MHz bandpass for the dMe star
AD~Leo at $1.4\,$GHz (20$\,$cm), but the spectrum here has opposite sense
($\alpha=+12$).  In both cases it seems difficult to
extrapolate the steep spectra over a wide range in frequency, but the
similarity may point to band-limited emission from both sources.



\subsection{Examinations of Detected Objects}
\label{sec:detect}
We now  examine the objects that we did detect in the
3-$\sigma$ error circle around the position of \gcrt.  We start with
the objects detected in $I$-band: as we discussed above, a
non-detection in $I$-band generally rules out objects of interest, at
least for spectral types earlier than L9 or so.

\begin{figure}
\plotone{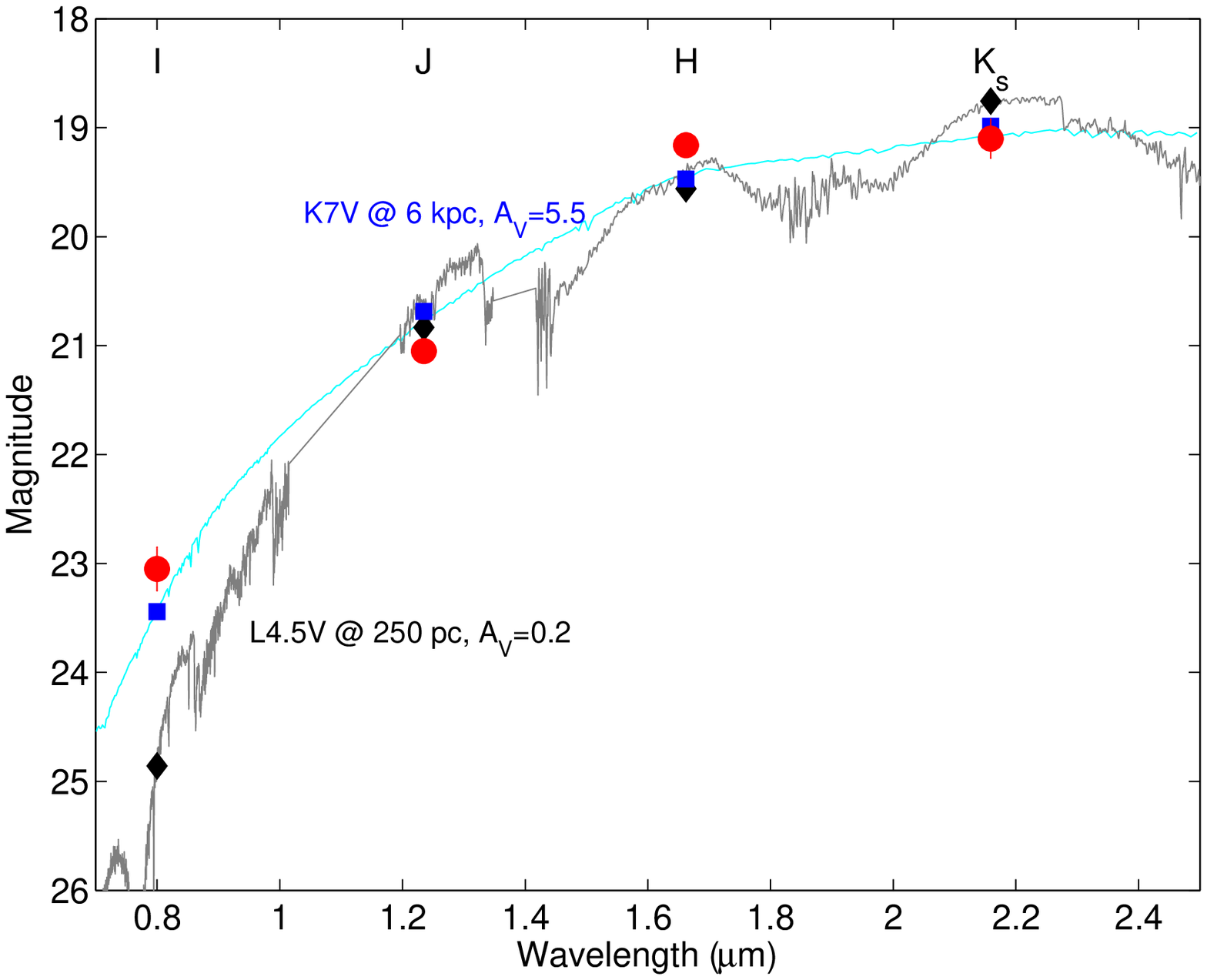}
\caption{Photometry of object C along with model fits.  We plot the
  $IJH\ks$ magnitude as a function of wavelength for object C (red
  circles), a K7V star at 6~kpc with $A_V=5.5\,$mag (blue squares;
  from \citealt{bcah98}, using the extinction laws of \citealt*{ccm89}
  and \citealt{imb+05} for the optical and infrared, respectively), and a L4.5V brown dwarf at $250\,$pc with
  $A_V=0.2\,$mag (black diamonds; the source 2MASS~J2224$-$01 from
  \citealt{dhv+02}).  We also plot spectra of similar comparison
  stars, where we have divided the spectra by a $10^4\,$K blackbody to
  get approximate Vega magnitudes: a $4300\,$K model atmosphere from
  \citet[][light blue]{ck04} and a L4.5V brown dwarf
  (2MASS~J1112256+354813) from \citet[][gray]{rbc+01}.  }
\label{fig:objC}
\end{figure}

In $I$-band, there are three objects that we detect: A, B, and C
(Tab.~\ref{tab:phot} and Fig.~\ref{fig:imglabel}).  We
plot them, along with the other objects from the field, in
Figure~\ref{fig:cmd}.  The first two objects are quite bright in the longer
wavelengths, but faint in $I$-band ($I-\ks=8.3$ and $I-J=5.8$ for A;
$I-\ks=11.2$ and $I-J=7.1$ for B).  In both of those cases, the colors
are generally too red for any sort of nearby, low-mass star or
ultracool dwarf, as we would not expect anything with $J-\ks>2$ or
$I-\ks>6$, as discussed above\footnote{We must be a little careful
with the $J$-band and especially $H$-band magnitudes for these stars,
as they are separated into several stars in $\ks$-band but not in the
bluer bands.  Therefore, the $J$ and $H$ magnitudes could actually be
lower limits.  This would tend to make the objects fainter in those
bands and hence \textit{redder}, though, so our conclusions are
reasonably secure.  The $I$-band magnitude also appears secure.}.
Instead, these objects are most likely distant, reddened giant stars
that are consistent with the bulk of the objects in the field
(see Fig.~\ref{fig:cmd}).

Object C is more interesting.  It is in the right color range
($J-\ks=4.0$, $I-J=2.0$) to be an ultracool dwarf.  However,
the colors are a bit contradictory, as can be seen from comparison
with the ultracool dwarfs plotted in Figure~\ref{fig:cmd}.  From the
$J-\ks$ color, we could guess a spectral type around late L, but the
$I-J$ color implies a much earlier object (early M).  The $H-\ks$
color is also bluer than expected for a late L object, but given the
blueward trend for T0 and later, it could be a T dwarf.  Some of these
discrepancies could be due to the difficulty in applying our $I$-band
calibration to a red object, but the $H-\ks$ and $J-\ks$ implications
differ as well.  Based on the $M_{\ks}(ST)$ relation from
\citet{vhl+04}, object C would have spectral type
\begin{equation}
ST_{C}\approx21.6-15.4\log_{10}\left(\frac{d}{100\,{\rm pc}}\right)
\end{equation}
(this relation is really only valid for L dwarfs).  So for the
nominal distance of $100\,$pc, we would expect a very late L/early T
object (again neglecting extinction).  

Fitting all of the objects in \citet[][]{dhv+02} and \citet{vhl+04} to
our data and restricting the fit to $JH\ks$, the best fits are for
spectral types around L5 (Fig.~\ref{fig:objC}) and distances around
200~pc (roughly consistent with the relation above).  There are
systematic deviations between the catalog objects and C, where they
underpredict the $J$- and $\ks$-band magnitudes (typically by about
0.3~mag) and overpredict the $H$-band magnitude (by a similar amount).
These differences are systematic in direction, but are not greatly
larger than the variation within the brown-dwarf sequence (typically
$<0.25$~mag; \citealt{dhv+02,vhl+04}).  Perhaps this variation can be
attributed to poor calibration of the other bands, although our checks
with 2MASS stars had much smaller residuals ($<0.1$~mag).  The catalog
photometry was generally in the Caltech (CIT) photometric system, and we converted it to the
2MASS system using the transformations from \citet{c01}.  For most
colors, the differences are small, but those transformations were not
necessarily appropriate for L or T dwarfs.  For the reddest stars
there could be deviations of $\sim 0.1$~mag between $J$-band in the
two systems, which would not really reconcile the differences
discussed above, but could help in some cases.  Perhaps age or
metallicity could be the culprits, although they are unlikely to be
significant for objects in the Galactic plane but outside of star
clusters.  A small amount of reddening helps the fit, but we cannot
accommodate very much reddening without being unphysical given the distance.

Much more significant is the difference in $I$-band magnitude for the
data from \citet{dhv+02}.  Fitting only the $JH\ks$ bands as discussed
above, we find that our $I$-band measurement is $\sim 1.5\,$mag brighter than the
catalog objects (such as the fit in Fig.~\ref{fig:objC}).  If we fit
all four bands, the quality of the fit gets considerably worse, with
deviations of $>0.5\,$mag common, and the best fits move to earlier
type stars (late M to L0) at greater distances.  These deviations are
far larger than what is seen among brown dwarfs, and do not seem like
they can be easily accommodated by reddening, metallicity, or
calibration issues.  It may be that there is a disk or other
complication that gives rise to different colors and explains the
radio emission, but without additional details this would be very
difficult to constrain.

However, object C is also quite compatible with being a main-sequence
star (Fig.~\ref{fig:objC}).  We get a roughly reasonable fit for all
four bands (residuals of 0.25~mag or so) using the main-sequence
photometry from a late K/early M star at $\sim 6\,$kpc with
$A_V\approx 5\,$mag \citep{allen,kurucz93,bcah98}.  Allowing for
differences in the reddening law and metallicity, such a
classification seems reasonable.



Beyond $I$-band, all of the objects detected at $J$ but not $I$ are
similar in colors to object B, with very red colors indicative of
distant, reddened giant stars.

\section{Constraints on Other Source Types}
\label{sec:other}

\subsection{A White Dwarf}
\citet{zg05} proposed a model for \gcrt\ where it was a ``white dwarf
pulsar'': a rotating, magnetized WD that, in analogy with traditional
pulsars, emits light due to conversion of rotational energy to
electromagnetic waves/particles through a strong magnetic field.  At
the wavelengths considered here, WDs generally have neutral colors
($J-\ks\approx 0$, etc.), since they are moderately well approximated
by warm-to-hot blackbodies with effective temperatures $>5000\,$K
\citep[e.g.,][]{bsw95,brl97,blr+01,whh+03,hb06c,hws+07} and the magnitudes we
are using are based on Vega, another hot blackbody.  While
approximate, this should be accurate to $\pm0.5\,$mag or so.  Based on
this, none of the objects detected in $I$-band is consistent with
being a WD.

\begin{figure}
\plotone{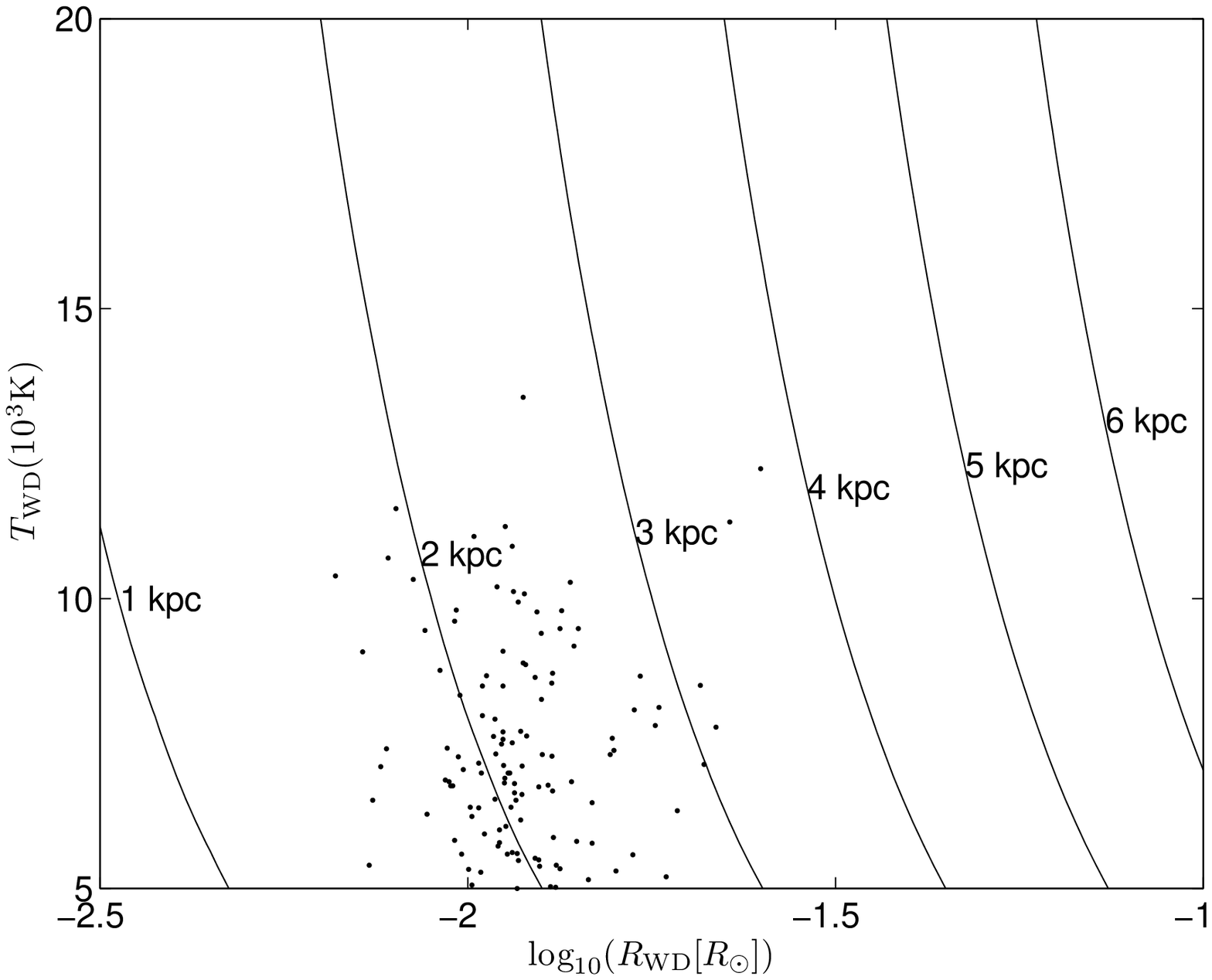}
\caption{Distance limits for any white dwarf in the error circle of
  \gcrt, based on the $I$-band non-detection.  We plot contours as a
  function of WD radius (in units of the Solar radius) and
  effective temperature.  The contours are labeled by the distance
  limits in kpc, assuming an average extinction of $1\,{\rm mag}\,{\rm
  kpc}^{-1}$.  The points are the cool WDs from \citet{blr+01}.}
\label{fig:wd}
\end{figure}

We next determine a distance limit inside of which there
are no WDs.  For the $I$-band magnitude of a WD with
radius $R$, effective temperature $T_{{\rm eff}}$, and distance $d$,
we have
\begin{eqnarray}
I_{\rm WD}(R,T_{\rm
  eff},d)&=&I_0-
 2.5\log_{10}\left(\left(\frac{R}{R_0}\right)^2\left(\frac{d}{d_0}\right)^{-2}\left(\frac{T_{\rm
  eff}}{T_{{\rm eff},0}}\right)\right)\nonumber \\
 && +A_I
\end{eqnarray}
where $I_0$ is the magnitude of some fiducial object with radius
$R_0$, effective temperature $T_{{\rm eff},0}$, and distance $d_0$,
and $A_I$ is the extinction at $I$-band (and the same for the other
bands).  This assumes that we are on the Rayleigh-Jeans portion of the
spectrum, which is largely true for $T_{\rm eff}>5000\,$K and
wavelengths redward of $I$-band, and takes care of the bolometric
corrections needed to transform a blackbody into observed bands.  For
the fiducial values, we use the average of the objects from the sample
of \citet{blr+01}: $R_0=0.012R_{\odot}$, $T_{{\rm eff},0}=7100\,$K,
$d_0=10\,$pc, and $(I_0,J_0,H_0,K_0)\approx 12.9$.  In reconstructing
the values from that sample, choice of a fiducial gives the other
sources to an accuracy of $\pm0.4\,$mag, comparable to the scatter of
the WD sample of \citet{whh+03} in the color-color plane (we are also
mixing WDs of different surface composition, but the differences at
wavelengths of interest are small).

We can then use these relations to limit WDs in the error
circle of \gcrt.  Since the limiting magnitude is highest for
$I$-band, that is the most constraining.  Here we need to take the
extinction into account more explicitly, since the distances are
$>500\,$pc and it will be significant.  Assuming an average extinction
of $1\,{\rm mag}\,{\rm kpc}^{-1}$, we find the distance limits plotted
in Figure~\ref{fig:wd}.  The limits are a function of WD
radius and effective temperature, but for typical temperatures
(5000--20,000$\,$K) and radii ($0.01R_{\odot}$) our $I$-band limit
implies no WD closer than $2\,$kpc.  We compare this lower limit with the
nominal distance upper limit of $0.8\,$kpc derived by \citet{zg05} for a WD
pulsar (with the constraint that the radio luminosity not exceed the
rate of rotational energy loss), and see that their basic model is not
consistent with our observations.  


There are also occasional WDs that are considerably cooler
(e.g., $\approx 3500\,$K for the source from \citealt{hoh+00}) and
which can have bizarre near-infrared colors ($I-J\approx 0$ but
$J-K\approx -1.4$ for the same object).  Such objects tend to be quite
blue in the near-IR, despite their cool temperatures.  The near-IR
flux is considerably less than one would predict from the optical
blackbody, with a deficit of more than one order of magnitude at
$2\,\um$.   $I$-band is better behaved, but the data are still
1.5$\,$mag fainter than our simplistic prediction above.  Even so, none of the
objects that we detected would be consistent with a cool WD since they
all have red $J-\ks$, and our $I$-band limit still constrains such an
object to be more distant than $\approx 1.4\,$kpc.

\subsection{A Star At the Galactic Center}
We can also see what limits our data can place on a possible stellar
object at the Galactic Center.  Here extinction is the largest
unknown.  We begin by assuming $A_V\approx 20$--30$\,$mag at a distance of
$8.5\,$kpc \citep[e.g.,][]{dsbb03}, but there could be a wide range of
possible extinctions (both in total column density and in variation
with wavelength; e.g., \citealt{nnk+06}; \citealt*{gbb06}; \citealt{gbmjf07}) due to local clouds.

With such a combination, our data are not very constraining.  We can
largely exclude supergiants based on the expected $\ks$ magnitudes:
the faintest (which are actually the hottest, such as B0I) would have
$\ks\approx 12$ (based on \citealt{allen}, although we note that
supergiants have a wide observed range in luminosity), while the brightest star we find in the error circle
has $\ks=14.2$ (object B).  So we would need 2 additional magnitudes of
extinction at $2\,\um$, or $\sim 20$ additional magnitudes at $V$-band
\citep[e.g.,][]{mathis90}, for a
total $A_V\approx 50\,$mag.  This is larger than we would expect,
although not completely unreasonable. It would, though, be rather
redder than what we observe for object B: we find $H-\ks=1.1$, but a
supergiant with $A_V=50\,$mag would have $H-\ks\approx 3$.  Variations
in the extinction law $A_H/A_{\ks}$ would have to be far larger than
observed \citep[e.g.,][]{nnk+06,gbmjf07} to agree with the data.
Later-type
supergiants would have even worse agreement.  They are brighter in the $\ks$-band,
so we would need additional extinction to get them to agree with our
observed magnitude.  But they are also redder, and with that
additional extinction they would be redder still, such that a K0I star
would have $H-\ks\approx 6$ (and would require $A_V\approx 80\,$mag!).

For other luminosity classes, though, we can say very little.  There
are certainly giant stars at the Galactic Center consistent with
objects in the error circle, and main sequence stars are even easier
to accommodate.  We cannot exclude any of the canonical mass donor
stars of Low Mass X-ray Binaries or Cataclysmic Variables (G--M main
sequence stars) at the distance of the Galactic Center.  For instance,
object B is generally consistent with a K2III star at 8.5$\,$kpc and
with an extinction of $A_V\approx 19\,$mag, although other
possibilities are certainly possible given the degeneracies between
reddening and spectral type, especially if one allows for a changing
extinction law.



\section{Discussion and Conclusions}
\label{sec:conc}
We have examined relatively deep images of the field of \gcrt\ taken
under good conditions (seeing better than $0\farcs5$) meant to search
for a possible nearby counterpart to the radio transient.  Such an
object would have to be at a distance $<70\,$pc in order to not
violate the $10^{12}\,$K brightness temperature limit on incoherent
sources.  Beyond that, we searched for nearby ultracool dwarfs at a range
of distances.

Examining data at wavelengths from $0.8\,\um$ to $2.1\,\um$, we find
that the field changes dramatically.  Only three objects are detected
at the shortest wavelength, compared to more than 20 at the longest
wavelength.  Two of the three sources detected at $0.8\,\um$ are
background giant stars.  The third (object C) is most likely a
late-type K star at a moderate distance ($\sim 6\,$kpc).  We cannot
entirely rule out the possibility that it is an object of type
$\sim$L5, but this is not  consistent with the $I$-band photometry
and would still be at a distance of $\sim 200\,$pc: further than the
$100\,$pc limit discussed above.  If it is of a late K type, object C
is unlikely to be the source of the radio emission. While isolated K
stars typically do not show strong radio flares, it is possible but
generally at fainter flux levels \citep{gudel02}.  Given the radio
luminosities typically seen ($\sim10^{23}\,{\rm erg}\,{\rm s}^{-1}$;
\citealt{gudel02}), it would be far too faint to be the source of the
radio emission from \gcrt.  However, the radio luminosities seen to
date for K stars may not include the brightest transient events.

Some K stars in binaries (e.g., RS~CVn binaries; \citealt{gudel02})
also show coherent radio flares.  Such objects have peak fluxes of
$<1\,$Jy, even at relatively low radio frequencies
\citep[e.g.,][]{vdodb94}, and for the distances implied for object C
(possibly underestimated if it is a binary, especially if it is
evolved like the K star in the prototypical RC~CVn binary HR~1099) it
would again be far too faint.

If we have not detected the counterpart, we use constraints from the
shorter wavelengths, where the field is much less crowded, to limit
the presence of any low-mass object.  For distances of $< 100\,$pc, we
can exclude stars earlier than T7 (with related limits of stars
earlier than T4 at $200\,$pc, M9 at $1.3\,$kpc, M5 at $2\,$kpc, etc.)
or so based on the $I$- and $J$-band photometry.

Our search is limited by the sample of comparison objects that we
used: nearby ultracool dwarfs with good photometry and (often)
astrometry.  We therefore cannot exclude the possibility that object C
is some peculiar type of ultracool dwarf, where the mis-match in the
$I$-band reflects some intrinsic difference of this object, but we
require additional constraints to be able to decide conclusively one
way or the other.  While the source is bright enough at $\ks$-band for
spectroscopy with large telescopes, $I$-band spectroscopy is likely to
prove most productive, as the source is much easier to isolate from
the nearby objects.  Additional photometry will suffer from the near
degeneracy between effective temperature and reddening seen in
Figure~\ref{fig:objC} and the difficulty of calibrating data where the
spectrum is dropping so sharply, but spectra should be able to
distinguish cleanly between the largely featureless spectrum of a
late-K/early-M star and the deep metal hydride absorption of an L
dwarf.  Our conclusions are also dependent on a lack of significant
variability: fluctuations of $<0.5\,$mag would not change things, but
larger variations might.  We see good agreement between the two sets
of near-IR data (from PANIC and NIRI).  However, those data were taken
within two nights of each other.  Variations over longer timescales,
such as the $\sim1\,$yr between the near-IR and optical observations,
or even short flares, could still be present, although we note that we
compared our data with photometry of the 3 sources (A, B, and D)
within the error circle that are visible in the UKIRT Infrared Deep Sky Survey
(UKIDSS) Galactic Plane Survey Data Release 3 (see
\citealt{lwa+07,lhl+07}) taken 2 years after our near-IR data and see
no sign of variability at levels $\gsim 0.3\,$mag; below that, it is
difficult to compare given the differences in calibration, seeing, and
photometric technique.  Additional monitoring could help address this
issue.

With the same data, we have attempted to constrain other source types
as possible counterparts to \gcrt.  We find that no white dwarf nearer
than $\approx 1.5\,$kpc could be present in the error circle, nor
could there be a supergiant star at the distance of the Galactic
Center.  However, both of those conclusions have the same limitations
as that discussed above.  Namely, we searched for objects that
resemble other, known objects.  If the radio emission from \gcrt\
really marks it as a unique object than the counterpart that we seek
could have dramatically different properties.  The presence of a
binary companion, accretion disk, or other anomaly could also
complicate things \citep[e.g.,][]{hws+07}, although most such
additions to the system would make it brighter and hence would still
largely be ruled out.

With all of the analysis above, we have excluded virtually any
white-dwarf or non-degenerate star as counterpart to \gcrt\ within
$100\,$pc, and many sources are excluded within $1\,$kpc.  We are
therefore left with the conclusion that \gcrt\ likely emits via a
coherent process.  What this means, though, is unclear.  Coherent
emission with brightness temperatures as high as $10^{15}\,$K (e.g.,
\citealt*{swr08}; \citealt{ob08}) and sometimes higher have been seen
from isolated stars and binaries, either relating to magnetic activity
from the stars or from interaction between the members of the binary
\citep{gudel02,osten08}.  The emission seen from \gcrt\ does not
really resemble the known radio emission from such sources, although
this may reflect the limitations of our observations.  As discussed
above, most sources have been observed to be significantly fainter
than \gcrt, but the comparison may be unfair since the behavior across
such a wide range of frequencies and timescales is not known.  If we
remove the flux/luminosity constraints, the emission from \gcrt\ is
actually not dissimilar to some ultracool dwarfs (such as the results
of \citealt{brr+05} and \citealt{had+06}).  Time-resolved analysis at
frequencies of $<500\,$MHz ($>60\,$cm) have been rare especially in
recent years \citep[e.g.,][]{vdodb94}, and we really do not know how
known radio sources behave there.  It seems likely that either \gcrt\
represents new low-frequency behavior from a known class of sources,
or that it is indeed the first member of a new class.
Searches for counterparts to \gcrt\ at other wavelengths (especially
X-rays, where the searches so far have not been very constraining;
\citealt{hlk+05}), along with further characterization of the radio
properties of both \gcrt\ and other transient sources, are required to help
elucidate its nature.

\acknowledgements We thank an anonymous referee for helpful comments,
and A.~Burgasser for useful discussions.  We also thank Jean-Rene Roy
for awarding us Director's Discretionary Time to obtain the NIRI
observations (under Gemini program GN-2005A-DD-13).  Partial support
for DLK was also provided by NASA through Hubble Fellowship grant
\#01207.01-A awarded by the Space Telescope Science Institute, which
is operated by the Association of Universities for Research in
Astronomy, Inc., for NASA, under contract NAS 5-26555.  SDH is
supported by funding from Research Corporation and SAO Chandra grants
GO6-7135F and GO6-7033B.  Basic research in astronomy at the Naval
Research Laboratory is supported by 6.1 base funding.  PyRAF is a
product of the Space Telescope Science Institute, which is operated by
AURA for NASA.  This research has made use of SAOImage DS9, developed
by the Smithsonian Astrophysical Observatory.

{\it Facilities:} \facility{GMRT}, \facility{Magellan:Baade (PANIC)},
\facility{Gemini:North (NIRI)}, \facility{Magellan:Clay (MagIC)}







\end{document}